\input epsf.sty
\documentclass{aa}

\begin{document}
 \thesaurus{03           
              (02.18.7;  
               11.01.2;  
	       11.17.3;  
	       11.19.1;  
               13.21.1;  
               13.25.2)}  
\title{Toy model of obscurational variability in active galactic nuclei} 


\author{A. Abrassart
\inst{1}
\and 
		B. Czerny \inst{2} 
}

\offprints{A. Abrassart}   

\institute{DAEC,Observatoire de Paris, Section de Meudon, F-92195 Meudon, 
France\\ 
email: noar@obspm.fr,suzy@obspm.fr,dumont@obspm.fr
\and
Copernicus Astronomical Center, Bartycka 18, 00-716 Warsaw, Poland\\ 
email: 
bcz@camk.edu.pl
}

\date{Received ...; accepted ...}

\maketitle

\begin{abstract}

We propose an explanation of the variability of AGN based on a cloud model of
accretion onto a black hole. These clouds, present at a distance of 10-100
$r_{Schw}$, possibly come from the disrupted innermost disk and they can
partially obscure the central X-ray source due to their extreme Compton
optical depth. We consider the implications of this scenario using a toy
model of AGN spectra and support the results with full radiative transfer
computations from the codes {\sc{titan}} and {\sc{noar}}. We show that 
small random rearrangements of the cloud distribution can 
happen on timescales of the order of $10^2 - 10^6$ s and may lead
to relatively high variability amplitude in the X-ray band if the mean covering
factor is large. The normalized variability amplitude in the UV band is
either the same as in the X-ray band, if the contribution from the dark sides of 
the clouds to the UV band is negligible, or it is smaller. The X-ray spectrum
should basically preserve its spectral shape even during the high amplitude 
variability, within the frame of our model. 

\end{abstract}

%

\section{Introduction}

Radio-quiet active galactic nuclei (AGN) are well known to be 
variable in the X-ray band 
since
early observations by EXOSAT (e.g. Lawrence et al. 1987, McHardy \& 
Czerny 1987, Green, McHardy \& Lehto 1993).  The combined
study of both the X-ray spectra and variability offer the most 
direct insight into the 
structure of
accretion flow onto the black hole which powers the AGN (for a review, see 
Mushotzky, Done 
\& Pounds 1993).

The variability trends have been extensively studied in various spectral
bands (e.g. Ulrich, 
Maraschi \& Urry 1997, Peterson et al. 
1998).
Generally, more luminous objects are less variable both in X-rays 
(Nandra et al. 1997 and the references therein) and in the optical/UV
 band
(Ptak et al. 1998, Giveon et al. 1999).

Multi-wavelength studies showed that this variability is 
surprisingly complex.  
In the X-ray
band, most of AGN vary but individual sources display various 
spectral trends 
with the brightening of the source (e.g. Ciliegi \& Maccacaro 1997, 
George et al. 
1998, Nandra et
al. 1997). The correlated variability between different energy 
bands is also difficult to
interpret (e.g. Nandra et al. 1998 for UV and X-ray connection in 
NGC 7469). 

The question appears to be whether this observed variability is entirely 
intrinsic, i.e. related  to
strongly non-stationary release of gravitational energy of the 
in-flowing matter, or is 
caused, at
least partially, by the effect of variable obscuration towards the 
nucleus. 

Partial covering models were popular mostly at the beginning of
spectral studies in the X-ray  band (e.g. Mushotzky et al. 1978, 
Matsuoka et al. 1986).
Recently, Seyfert 1 galaxies and QSOs are modeled through an accretion
disk, with X-ray emission coming either from the disk corona or
from the disrupted innermost part of the disk (e.g. Loska \& Czerny 1997
and the references therein).

However, occasionally, the problem was revived. An eclipse by a 
cloud was successfully 
considered 
as a model for the faint state of MCG-6-30-15 (McKernan \& Yaqoob 1998, 
Weaver \& 
Yaqoob 1998) and it may be a possible cause of the $K_{\alpha}$ line
profile variability in NGC 3516 (Nandra et al. 1999). In the case
of the variability of Narrow Line Seyfert 1 galaxies (e.g. Boller et al. 1997 for IRAS 13224-3809; Brandt et al. 1999
for PHL 1092) it was 
suggested
that the observed huge amplitudes  
are hard to explain 
directly
through a variable energy output, and instead, partial covering and/or relativistic 
erratic beaming may be necessary. It is interesting to note that the partial 
covering mechanism may also apply to similarly variable galactic sources (Brandt et al. 1996 for Cir X-1). Obscuration events are also sometimes 
invoked to explain the variations of the Broad Emission Lines in Seyfert 
galaxies - for example, a spectacular change of morphological type
from Seyfert 2 to Seyfert 1 of the galaxy NGC 7582 (Aretxaga et al. 1999) 
was interpreted
as due to an obscuration event by a distant cloud (Xue et al. 1998). 
Finally, there is
a large amount of partially ionized material lying along the line of
sight to the nucleus which manifests itself as a warm absorber 
(cf. Reynolds 1997). 
Variability in absorption edges can either be due to a variation of the 
ionisation state of the absorber or to obscuration by clouds passing through 
the line of sight.

Although there are now strong suggestions that the accretion flow 
proceeds 
predominantly through  a
disk the physics of the innermost part of the flow is highly 
uncertain. As the inner 
regions of
a standard Shakura-Sunayev (1973) disk are thermally and viscously 
unstable in an AGN 
owing
to the large radiation pressure, the disk is frequently supposed to 
be disrupted. It may 
form
a hot, optically thin, quasi spherical 
ADAF (advection-dominated accretion flow; Narayan \& Yi 1994, and many 
subsequent papers),
or it may proceed in the form of thick clouds (Collin-Souffrin \& al 
1996, hereafter 
referred as
Paper I). Such clouds might be optically thick, unlike 
clouds optically thin for 
electron
scattering which form spontaneously as a result of thermal 
instability in X-ray irradiated 
cold gas
and coexist in equilibrium with the surrounding hot optically thin 
gas (Krolik 1998,
Torricelli-Ciamponi \& Courvoisier 1998).  Such clouds provide significant 
mass flux and they are not easily destroyed (see Sect. 4 in Paper I); they
may even condensate under favorable conditions but the criterium for 
evaporation or condensation depends on the assumed heating 
mechanism (see R\' o\. za\' nska \& Czerny 1999 and the references therein). 
In another model 
(Celotti, Fabian \& 
Ress, 1992,
Sivron \& Tsuruta 1993, Kuncic, Celotti \& Rees 1996) very dense 
blobs confined by 
magnetic
pressure, optically thin to scattering and thick to free-free 
absorption, form in a 
spherical
relativistic flow.
 
Within the frame of the cloud scenario, obscuration events are 
expected. If clouds have a
column density in excess of $\sim 10^{25}$ cm$^{-2}$,  the  
obscuration events would be observed as 
variability,
whatever the size of the clouds (smaller or larger than that of the 
central source).

In the present paper we discuss the possibility of explaining, at 
least partially, the 
variability
phenomenon by variable obscuration. We consider the case of random 
variability due to 
the statistical
dispersion in location of clouds along the line of sight for a 
constant covering factor. 
We provide simple
analytical estimates of the mean spectral properties, timescales and 
variability amplitude of 
AGN,
and we support them with computations of radiative transfer done 
with the use of the codes {\sc{titan}} (Dumont, Abrassart \& Collin 1999) 
and {\sc{noar}} (Abrassart 2000).


\section{Model}

\subsection{Cloud model scenario}

\begin{figure}
\epsfxsize=8.8cm \epsfbox{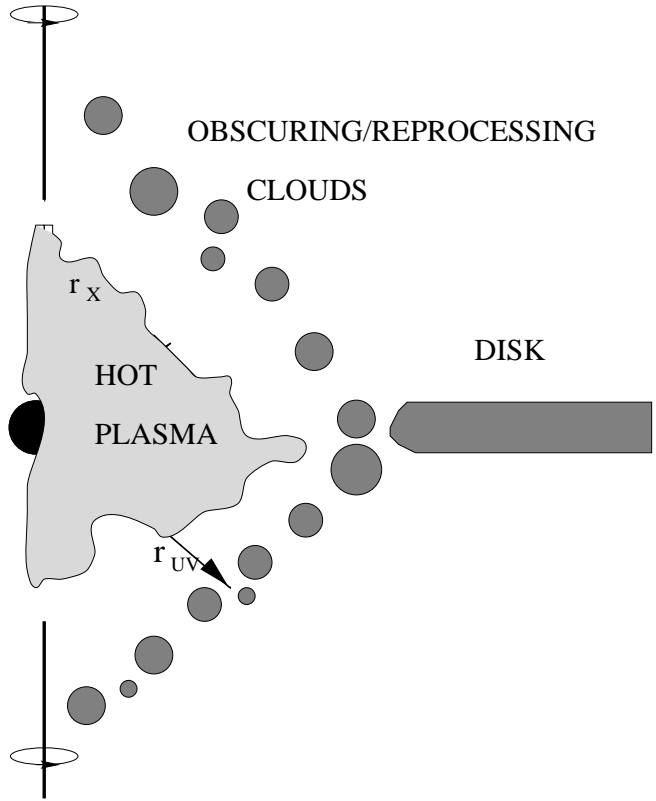}
\caption{A fantasy view of the cross-section through the central 
region of the accretion flow 
into a
black hole. In the paper, we idealize this geometry assuming 
a spherically symmetric 
distribution of the
clouds at a characteristic radius and a spherical shape for the hot 
central plasma.} 
\label{fig:pict}
\end{figure}

Accretion flow onto a black hole proceeds most probably down to a
hundred Schwarzschild radiiin the 
form of 
an accretion disk. 
Closer in, the relatively 
cool accretion disk is 
disrupted
and replaced with a mixture of still cold optically thick clumps 
and hot gas which is 
responsible for
hard the X-ray emission. The cold clumps may not be constrained to the 
equatorial plane but, 
under the
influence of radiation pressure, may have a quasi-spherical 
distribution. We envision this 
general
picture in Fig.~\ref{fig:pict} (see also Fig. 1 in Karas et al. 1999). 
Such a scenario offers an 
interesting possibility for 
explaining the
observed spectra of AGN. Although other geometries of the innermost 
part of accretion 
flow are not
excluded, we concentrate on exploring this particular model.

The mechanism of disk disruption is unknown and therefore the 
dynamics of the cloud 
formation
process cannot be described. The formation of the hot medium is also 
not well understood, 
although it
must proceed in some way through cloud evaporation. Therefore we 
have to resort to a 
phenomenological
parameterization of the cloud distribution and the hot gas geometry.
A realistic approach to the cloud model would require a large number 
of arbitrary  
parameters
reproducing the radial structure and the departure from spherical 
symmetry. However, the
most essential
properties of the model can be studied within the frame of a much 
simpler model.

In the present paper we follow the geometry adopted by Czerny \& 
Dumont (1998)  for 
their model C
(see their Fig. 2).

We assume that the hot medium forms a central spherical cloud of 
radius $r_X$.  It is 
characterized
by its Thomson optical depth, $\tau$, and by its electron 
temperature, $T_e$. Those two 
parameters
uniquely determine the Compton amplification factor $A(\tau, T_e)$ 
of the hot medium.

The cold clouds are located at a typical distance $r_{UV}$ from the 
center  and their 
distribution
is characterized by the covering factor $\Omega /4 \pi$. They 
reprocess the X-ray 
emission from
the hot medium. A fraction of the incident X-ray photons is backscattered, 
but most of 
the photons
are absorbed. This absorbed radiation is mostly reemitted in UV by the 
bright side of the clouds, 
but a
fraction is leaking through the dark side. The photons reemitted by 
the bright side provide 
seed
photons for Comptonization by the central hot medium. 

The bremsstrahlung emission from the central comptonizing cloud 
(CCC) does not  
significantly
contribute to the energy budget, for all the parameters we 
considered. In this model the 
prime
movers of the engine are the X-rays, in the sense that they are 
produced in a region (of 
radius
$r_X$) close to the black hole, and they drive the reprocessed UV 
emission further away. 
But the intrinsic
emission of the hot plasma being small, it basically acts as a 
reservoir of energy, used to
upscatter a fraction of the reprocessed UV photons. As the thick 
clouds can be highly 
reflective in
the UV, this ``recycling'' of photons can be very efficient when 
$\Omega /4 \pi $ is large. 
The
efficiency of the subsequent production of X-rays is also high if 
the scattering probability 
of UV
photons, determined by the optical depth of the hot plasma $\tau$ and 
the relative 
geometrical
cross-section $(r_X/r_{UV})^2$, is not too low. This geometry may thus 
imply a 
stronger coupling
between the two emission regions than the alternative disk/corona 
configuration, where 
about half of
the comptonized power escapes directly.

\subsection{Radiative transfer computations} 

An accurate description of the radiative coupling between the hot 
plasma and the 
relatively cold
clouds requires complex computations of radiation transfer. In 
particular, the emission 
from the 
bright side of the clouds strongly depends on the ionization state of 
the gas.

We calculate the emission from the bright and dark sides of the clouds 
in the optical/UV range 
using the 
radiative transfer code {\sc{titan}} of Dumont, Abrassart \& Collin (1999) 
for Compton thick
media. The same code was used by Czerny \& Dumont (1998) over the 
entire optical/X-ray 
range. However,
in the present version we pay more attention to an accurate 
description of the hard X-ray 
transfer
within the hot plasma and within the surface layers of clouds, 
therefore we use the
Monte Carlo code {\sc{noar}} of Abrassart (2000).

In order to obtain a single spectrum model, both codes - {\sc{titan}}
and {\sc{noar}} - have to be 
used in a form of 
iterative
coupling, as described in detail by Abrassart (2000). The method is 
very time consuming, 
so for
the purpose of the present studies of variability, we develop a  
simple analytical toy 
model which allows estimates of the observed trends in 
a broad range of parameters. Therefore, we use 
the numerical results as a test 
of the toy
model and as a source of information about physically reasonable 
mean quantities such as the frequency-averaged albedo, the fraction of radiation lost by the dark 
sides of the clouds, etc.

\subsection{Toy model}
\label{subsectoymod}
\begin{figure}
\epsfxsize = 8.8cm \epsfbox{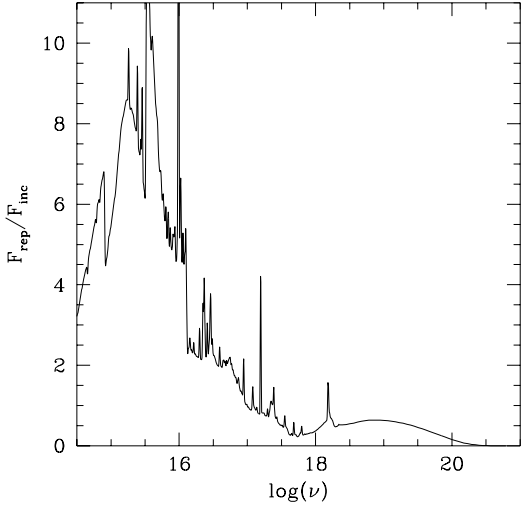} 
\caption{The ratio of the broad band reprocessed spectrum (i.e. reflected and 
reemitted) to the incident 
power law spectrum in the plane parallel 
geometry
for the value of the ionization parameter $\xi=300$ (see text).} 
\label{figalb} \end{figure}

In numerical simulations both the scattered and the reemitted 
component form a complex 
reflected 
spectrum with a shape mostly determined by the ionization parameter 
\begin{equation}
\xi =4 \pi F_{inc}/n,
\label{eq:xi} 
\end{equation}
where $F_{inc}$ is
the incident radiation flux and $n$ is the number density of a 
cloud, although the shape of the observed spectrum is even more influenced
by the value of the cloud covering factor. In 
Fig.~\ref{figalb} we
show the ratio of the radiation reflected/reemitted by the bright 
side of the irradiated cloud 
for a
value of the ionization parameter $\xi$ equal 300 and a power 
law spectrum proportional to $\nu^{-1}$  (the ratio depends little 
on the 
density and
column density, provided the latter is larger than 10$^{25}$ 
cm$^{-2}$, for a range 
of density
between 10$^{10}$ and 10$^{15}$ cm$^{-3}$). As we see, the true 
absorption is 
very low below the
Lyman discontinuity, even for such a relatively low $\xi$, the 
reflectivity is very close to 
unity,
and the absorbed and thermalized X-ray flux emitted in the UV creates an 
excess seen in 
UV/soft X-rays. 

Therefore, due to reprocessing by cold clouds, the initial almost 
power law distribution of 
photons 
produced by Comptonization is replaced by a still broad hard X-ray 
component and a much 
more peaked UV/soft
X-ray component. 
The X-ray component is produced by subsequent scattering of
photons within the hot cloud, thus forming 
basically a 
power law spectral shape with a high energy extension and a
slope determined by the 
properties
of the hot cloud ($\tau, T_e$). Cloud emission is closer to a black 
body and the maximum of the spectrum is mostly determined  by the temperatures
of the dark and bright sides of the clouds.

Here, for the purpose of an analytical analysis, we simplify this 
distribution. 

The entire optical/UV/X-ray range is schematized by defining two 
representative energy 
ranges: soft
(UV) or hard (X-ray), with mean energies $E_{UV}$ and $E_X$. The 
exact values of 
those energies are
not essential as they do not enter as model parameters through 
conservation laws (see 
Appendix A). 

We reduce the description of radiative transfer to coefficients 
which determine the 
efficiency of 
the change of an X-ray photon into a UV-photon, and the reverse. 

We reduce the photon-frequency-dependent albedo, describing the 
efficiency of reflection 
of X-ray
photons, to some energy-integrated value, $a$. We introduce a 
frequency-averaged fraction 
of the X-ray
luminosity, $\beta_d$, which leaks through the dark side of the clouds 
in the form of UV 
emission
instead of being reemitted by the bright side. 

We also simplify the description of Comptonization by using a mean 
Compton 
amplification factor 
independent of the energy of the input soft photon. 


Variability properties are related to the mean spectral shape of an 
AGN. Therefore we 
have to
introduce a relation between the cloud distribution and the observed 
spectral shape before
discussing the variability itself.

\subsubsection{Mean properties of the cloud distribution} 

We can formulate the conditions for the stationary situation, in 
which the escape of the 
photons 
from the system should be compensated by the production of new 
photons at the expense 
of the energy
flux supplied to the hot cloud. Noting $C=\Omega /4 \pi$ and 
$\gamma=\tau/(1+\tau)(r_X/r_{UV})^2$,
the fate of a UV photon at every travel across the region is 
determined by the following
probabilities: 

\begin{equation} (1 - \gamma)(1- C) = P_{esc}^{UV} \label{puvesc}
\end{equation}
\begin{equation}
(1 - \gamma) C = P_{refl}^{UV}
\label{puvrefl}
\end{equation}
\begin{equation}
\gamma=P_{ups}^{UV},
\end{equation}
where $P_{esc}^{UV}$ is the probability of escape from the system, 
$P_{refl}^{UV}$ is the probability of reflection by the clouds and
$ P_{ups}^{UV}$ describes the probability of conversion into an X-ray
photon due to the Compton upscattering. 

For an X-ray photon we have the following probabilities: 
\begin{equation}
1 - C = P_{esc}^X
\label{pxesc}
\end{equation}
\begin{equation}
a C=P_{refl}^X
\label{pxrefl}
\end{equation}
\begin{equation}
(1-\beta_d)(1-a) C=P_{abs}^X
\label{pxabs}
\end{equation}
\begin{equation}
(1-a)C \beta_d=P^{UV}_{dark},
\label{puvdark}
\end{equation}
where $P_{esc}^X$ is the escape probability, $P_{refl}^X$ is the probability
of reflection by the clouds, $P_{abs}^X$ describes the probability for an X-ray
photon to be absorbed by the clouds and reemitted in the form of UV radiation
through the bright side, while $P^{UV}_{dark}$ describes the probability of
absorption of an X-ray photon and subsequent reemission in the form of UV 
radiation through the dark side.

We have to compensate for the loss of UV photons with new 
UV photons created in a sequence of events: upscattering a fraction 
of UV photons by the hot plasma and creating X-rays, absorption of X-rays by the cold
clouds, and reemision of this energy in the form of UV photons 
(see Appendix A): 

\begin{equation} {(P_{esc}^{UV}
+ P_{ups}^{UV} )(P_{esc}^X + P_{abs}^X) \over P_{ups}^{UV}P_{abs}^X } = A(\tau, T_e).
\label{stat} \end{equation}

The stationarity can be clearly achieved only if $A$ is greater than 
unity. 

In the present formulation of the model we assumed that the emission 
from the dark sides of 
the clouds
is provided by X-ray photons. We could instead assume that it is mostly the 
leaking of UV 
photons
diffusing through an optically thick cloud which powers the dark 
side emission. This 
would mean
adding the factor $(1 - \beta_d)$ to the left side of Eq.~\ref{puvrefl}, 
adding a 
new equation to the set
formulated for a UV photon: $(1 - \gamma)C\beta_d=P_{dark}^{UV}$, while
putting $\beta_d=0$ in Eq.~\ref{puvdark}, and dropping 
Eq.~\ref{pxabs} in the set
formulated for an X-ray photon.
These two 
approaches are almost
equivalent unless $\gamma$ is very close to 1, corresponding to a 
change of all UV 
photons into
X-rays. 

The ratio of the intrinsic bolometric luminosities of these two spectral 
components inside the 
region is
given by (see Appendix A) 

\begin{equation}
\left({L_X \over L_{UV}}\right)_{int} ={P_{esc}^{UV} + P_{ups}^{UV} 
\over 
P_{abs}^X}
= { A \gamma \over 1 - aC}. \end{equation}
This equation shows that for a given system (specified by C, a, and $\gamma$),
 $L_X \over L_{UV}$ is upper bounded by the limited heat supply to the central comptonizing cloud. 

\subsubsection{Mean properties of the observed spectrum} 

The observed spectrum is not identical to the spectrum inside the 
production region
since the escape probability for UV and X-ray radiation differs by a 
factor of $(1 - 
\gamma)$
(compare Eqs. \ref{puvesc} and \ref{pxesc}), and X-rays transformed 
into dark 
emission additionally
change the relative proportions.

The observed ratio of the bolometric luminosities in the two 
components is therefore 
given by:

\begin{equation}
\left({L_X \over L_{UV}}\right)_{obs} ={\left({L_X \over 
L_{UV}}\right)_{int} (1 - 
C) \over  (1 -
C)(1 - \gamma) + \left({L_X \over L_{UV}}\right)_{int} \beta_d (1 -
a) C}. \label{lrat3}
\end{equation}

This formula simplifies considerably
if the contribution of the dark sides of the clouds is neglected:  

\begin{equation}
\left({L_X \over L_{UV}}\right)_{obs} =\left({L_X \over 
L_{UV}}\right)_{int}{1 
\over (1 - \gamma)} 
\label{lrat2} \end{equation}

This is true
if the clouds are very thick, as then their dark side temperature is low 
and the  radiation flux 
peaks in the
optical band.

It is worth noting that the observed luminosity ratio can be expressed 
{\bf independently} of $\gamma$ or $\beta_{d}$:
\begin{equation}
\left({L_X \over L_{UV}}\right)_{obs} =\frac{A\,\left( -1 + C \right) }{1 + a\,\left( A - 1 \right) \,C - A\,C}.
\label{eq:lrat5}
\end{equation}


\subsection{Model of stochastic variability}

We can consider two variability mechanisms expected within the frame 
of the cloud 
model and of the 
geometry of the inner flow. 

Random variability at
a certain level is always expected since the cloud distribution is 
assumed to be uniform 
(spherically symmetric) in a statistical sense. Even without any 
systematic evolutionary 
changes
of the covering factor $C$, our view of the nucleus will undergo 
variations due to the 
instantaneous
covering factor of the side of the distribution facing the observer.

Systematical trends may also show up if there is a systematic 
evolutionary change in the 
mean 
covering factor, $C$. 

\subsubsection{Random variability amplitude}
\label{sect:randam}

The most basic parameter of the variability is its amplitude at a 
given wavelength. In order 
to 
reproduce it within the frame of the cloud model we have to assume 
that clouds can have 
some random
velocities which do not change the mean covering factor but change 
randomly the 
number of clouds
which occupy the hemisphere just facing us. 

If we have $N$ clouds, half of them on average occupy the front 
hemisphere and the 
dispersion 
around that mean value would be given by

 \begin{equation} \delta (N/2) = \sqrt {N/2}
\end{equation}
so the covering factor as seen by us also displays variations with 
an amplitude 

\begin{equation}
\delta C = C \sqrt {2/N}
\end{equation}

Therefore the amplitude of variability of the observed X-ray 
emission is given by

\begin{equation}
\left({\delta L_X \over L_X}\right)_{obs} = { C \sqrt {2/N} \over 1 
- C}. \label{xampl}
\end{equation}

High X-ray variability is achieved if the covering factor is high 
and the number of clouds 
is not
too  large.

The UV variability depends again on the optical depth and the 
contribution of the dark 
sides 
predominantly to the UV or optical band. 

If dark sides of the clouds are too cool to radiate in the UV, the 
variability amplitude is given 
by the
same formula as for X-rays 

\begin{equation} \left({\delta L_{UV} \over L_{UV}}\right)_{obs} = { 
C
\sqrt {2/N} \over 1 - C}. \label{uvampl1} \end{equation}

However, if the dark sides contribute to the UV we have 

\begin{equation}
\left({\delta L_{UV} \over L_{UV}}\right)_{obs} = {C \sqrt {2/N} [(1 
-\gamma)(1 - 
aC) - \gamma A 
(1 -a) \beta_d] \over (1 - \gamma)(1 - C)(1 - aC) + \gamma A (1 -a)C 
\beta_d} 
\label{uvampl2}
\end{equation}
which reduces the UV relative amplitude. 

Such normalized variability amplitudes depend on the number of clouds, $N$,
constituting the UV emitting region. This means that an additional free 
parameter is involved.

The normalized variability amplitudes can be determined observationally. 
Studies
of variability provide us either with the rms value in a given spectral band,
or just with a typical value of the variability factor. Quoted rms 
values
can be directly identified with our normalized amplitude. If instead only
a variability factor is given, we can understand it as statistical 
variations at
2 standard deviation level. 
This means that we actually observe variability of a factor ${\cal 
A}$ when the
luminosity changes from $L_X - 2 \delta L_X$ to $L_X + 2 \delta L_X$. 

Therefore this variability factor can be
expressed as

\begin{equation}
{\cal A} = L_X (max)/L_X(min) ={1 + 2\left({\delta L_X \over L_X }\right)_{obs} \over 1 - 2\left({\delta 
L_{X} \over 
L_{X}}\right)_{obs}} . 
\end{equation}

We can also define the ratio $R$ as the ratio of the normalized 
variability
amplitudes in the X-ray band and in the UV band:

\begin{equation}
R={\left({\delta L_X \over L_X }\right)_{obs} \over \left({\delta 
L_{UV} \over 
L_{UV}}\right)_{obs}}. 
\label{eq:R}
\end{equation}

Such a ratio does not depend on the number of clouds, $N$, constituting the
UV emitting medium. Therefore this ratio is fully determined by the
parameters of our toy model.

\subsubsection{Random variability timescales}

The random variability we consider results from the clouds passing 
through our line of sight to the central source. On the one hand, each passing
cloud produces an eclipse phenomenon. On the other hand, the dispersion 
in the cloud velocities, due to even a small dispersion of the distances from
the gravity center, results in variations of the cloud distribution,
as described in Sect.~\ref{sect:randam}.

Let us estimate a representative duration $t_{dip}$ of 
an obscuration 
event. This  is
determined by the size and velocity of a cloud as well as the size
of the X-ray source and the entire region involved. The cloud velocity 
is of the order of the 
keplerian
velocity so it is determined by the mass of the black hole (or its 
gravitational radius) and  $r_{UV}$. A
typical size of a 
cloud is given simply by the number of clouds, the radius $r_{UV}$ and the 
covering factor, if there is at most one cloud in the line of sight, i.e. clouds
are not overlapping.

The mean number of clouds in a line of sight is of the order of
unity (cf Appendix B) so we get the typical cloud size:

\begin{equation} 
r_{cl}^2\sim {C \over N} r_{UV}^2. \label{clsize}
\end{equation}

Here we have ignored numerical factors of the order of $\pi$, since this
expression, and the following formulae, serve only as an order
of magnitude estimate.

Since the clouds are ten or more times smaller than the region involved,
the fastest variation corresponds to an ingress or an egress from
a single eclipse

\begin{equation}
 t_{min}\sim {r_{Schw} \over c}\sqrt{{4 C\over N} 
({r_{UV}\over R_{Schw}})^3} \label{3kepler}. \end{equation}

Typical  random rearrangement of the cloud distribution will proceed
on a dynamical timescale connected with Keplerian motion

\begin{equation} t_{d}\sim {r_{Schw} \over c}\sqrt{
({r_{UV}\over r_{Schw}})^3} \label{dynamic}. \end{equation}
 
If the cloud distribution covers a range of radii, the longest
timescale involved will be given by $t_d$ at the outer edge
of the cloud distribution.

\subsubsection{Variations of the covering factor in the toy model} 

This kind of variability is more complex to study since variations 
in the covering factor 
lead not only directly to variations of $L_X$ and $L_{UV}$ but 
also
result in a change of the mean number of 
clouds along the line of sight, and, under an assumed constant total 
luminosity, in a change of the 
ionization state of the clouds, i.e. albedo. Such trends can only be 
studied by solving the 
radiative transfer equation for a sequence of models, which we postpone to a 
future paper.

\subsection{Relation between toy model parameters and the ionization parameter}
\label{parxi}
The spectral features in X-ray band allow us to have an insight into the ionization 
state of the reprocessing medium. This ionization state is mostly determined
by the value of the ionization parameter $\xi$. Since we frequently have some direct
estimates of this parameter, it is interesting to relate this parameter to the
parameters of our toy model. It will allow for simple estimates of those
parameters through the quantities which can be estimated on the basis of observational
data. That way, we can judge to some extent the applicability of the 
model.

For a bolometric luminosity of an object, $L$, the incident flux 
determining
the ionization parameter, $\xi$, 
(Eq.\ref{eq:xi}) in the case of the quasi-spherical distribution 
of clouds is: 

\begin{equation}
F_{inc} = {L \over r_{UV}^2 },
\end{equation}
where $n$ is the number density of the clouds and $r_{UV}$ is 
the representative distance of a cloud from the black hole. 

The most convenient form of  relating the number of clouds to the ionization parameter
and other observables is through the 
explicit 
dependence on the
 covering factor and the column density of the clouds, $N_H$. We can 
derive the
appropriate formula knowing that the typical size of the cloud 
$r_{cl}$ is related to $N$, 
the
covering factor and the characteristic radius of the cloud 
distribution (see Eq.~\ref{clsize}).

The obtained relation

\begin{equation}
N \sim  L^2 C N_H^{-2} r_{UV}^{-2} \xi ^{-2}. 
\label{eq:num}
\end{equation}
indicates that the number of
clouds scales with the square of the luminosity to the Eddington 
luminosity ratio, 
$R_{Edd}$, if
$N_H$, $\xi$ and  
$r_{UV}$ (expressed in
Schwarzschild radii) are similar in all objects. It means that high 
luminosity, low covering 
factor objects should be 
the least
variable. If this estimate of the number of clouds is inserted into 
Eqs.~\ref{uvampl1} and 
\ref{xampl}
we see that we do not expect a direct scaling with the central mass 
of the black hole or 
bolometric
luminosity. The very weak trends between the UV variability and the 
bolometric 
luminosity observed
in the data (Paltani \& Courvoisier 1997, Hook et al. 1994) might be 
caused by some 
secondary
coupling between the mass of the black hole and the model 
parameters. 

The size and the number of clouds is constrained by the dimension of 
the emission
region, as the volume filling factor has to be smaller than unity:

\begin{equation}
{N r_{cl}^3\over r_{UV}^3}\ll 1, \label{eq:filling-factor}
\end{equation} 

or:

\begin{equation}
N_H \xi\ r_{UV}\ L^{-1}\ll 1. \label{eq:filling-factor-bis}
\end{equation}

\medskip

One can now illustrate the properties of the cloud model with some numbers. 

Expressing the mass
of the black hole in 10$^8$M$_{\odot}$, $R_{Edd}$ in 10$^{-1}$, 
$N_H$ in 
10$^{26}$ cm$^{-2}$,
$\xi$ in 10$^3$ and $r_{UV}$ in 10 $r_{Schw}$, one gets, using  Eqs. 
\ref{clsize} and
\ref{eq:num}:

\begin{equation}
r_{cl}\sim 2\times 10^{12} N_{26} \xi_3\left({r_{UV}\over 
10 r_{Schw}}\right)^2 
M_8
\left({R_{Edd}\over 0.1}\right)^{-1}{\rm cm}
 \label{eq:clsize-bis}
\end{equation} 

\begin{equation}
n_H\sim 5\times 10^{13} \xi_3^{-1}\left({r_{UV}\over 
10 r_{Schw}}\right)^{-2} M_8^{-1}
{R_{Edd}\over 0.1}\ {\rm cm}^{-3} \label{eq:density}
\end{equation} 

\begin{equation}
N\sim 10^4 C N_{26}^{-2} \xi_3^{-2}\left({r_{UV}\over 
10r_{Schw}}\right)^{-2}\left({R_{Edd}\over  0.1}\right)^2,
\label{eq:eq:num-bis} \end{equation} 
and the condition on the filling factor:

\begin{equation}
 10^{-3} C  N_{26} \xi_3{r_{UV}\over 10r_{Schw}}\left({R_{Edd}\over 
0.1}\right)^{-
1}\ll 1,
\label{eq:filling-factor-ter} \end{equation} 

Clouds with such properties could form for instance 
as broken pieces of 
the inner
accretion disk. 


Using Eqs. \ref{eq:num} and \ref{xampl} the X-ray variability 
amplitude is:

\begin{equation}
\left({\delta L_X \over L_X}\right)_{obs} \sim {\sqrt C\over (1-C)} L^{-
1}N_Hr_{UV}\xi 
\label{xampl-bis}
\end{equation}

It is maximum for a covering factor equal to 0.5, corresponding to 
two clouds along the line 
of sight, on average.

Using the same notation as in Sect.~\ref{parxi}, the X-ray variability amplitude is:

\begin{eqnarray}
\left({\delta L_X \over L_X}\right)_{obs} &\sim & \sqrt{2} \times 10^{-2}
{ \sqrt C\over (1-C)} N_{26} 
\xi_3
\\
\nonumber
&\ &\left({R_{Edd}\over 0.1}\right)^{-1}{r_{UV}\over 10r_{Schw}}. 
\label{xampl-ter}
\end{eqnarray}
and the time scale of the variability, using Eq. \ref{3kepler}:

\begin{equation} t_{dip}\sim 10^2 \ N_{26}\  
\xi_3\left({R_{Edd}\over 
0.1}\right)^{-1}M_8 \ {\rm sec}\label{3kepler-bis} \end{equation}

We see that a large variability amplitude in a very small time 
scale is easily obtained in this model. 

We cannot expect, however, that the overall variability properties 
of AGN may be 
explained
by eclipsing clouds all being at the same distance and having the 
same radius. Actually, 
any more
realistic picture would require clouds to occupy a range of radii.


\section{Results}

The complete toy model of the stationary distribution of clouds 
depends on four 
parameters: the covering factor, $C$, the probability of 
upscattering a $UV$ photon into an
X-ray photon, $\gamma$, the X-ray albedo of the bright side of the
clouds, $a$, and the 
fraction of the X-ray radiation leaking through the dark side of the 
clouds in the form of 
$UV$ radiation, $\beta_d$. These four values determine the Compton 
amplification
factor of the hot plasma, $A$, through the conservation law given by 
Eq.~\ref{stat}. In full radiative transfer numerical models the number of free 
parameters is the same: $C$, $r_X/r_{UV}$, $r_X$, and the hot plasma 
parameters 
$\tau$ and $T_e$. Fortunately the number density and the column density of 
the cloud system 
play a small role. However the freedom in modeling is 
considerable. Therefore, in 
order to study the cases which are of direct interest to observed 
AGN, we consider first in 
some detail the mean Seyfert spectrum in order to fix some of these 
parameters to be used in most of our further 
considerations.

\subsection{Toy model of the mean properties of Seyfert 1 galaxies} 
\label{sect_mean}

\begin{figure}
\epsfxsize = 8.8cm \epsfbox{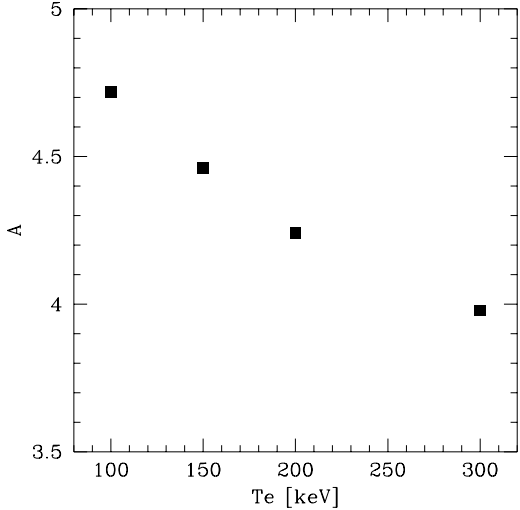} 
\caption{The dependence of the Compton amplification factor on the 
hot plasma
temperature, $T_e$, for the optical depth of the plasma adjusted for 
each temperature to fix 
the hard
X-ray slope index at 0.9. The soft photon temperature is 50 eV.} 
\label{figamplif} 
\end{figure}

The broad band spectra of Seyfert 1 galaxies are relatively well 
known. 
Most objects have an intrinsic hard X-ray slope of about $\sim 0.9$ 
if the spectra are 
corrected for the reflection component (Nandra \& Pounds 1994). 
However, determination of the shape of the high energy tail of the spectrum
of a single object 
poses a considerable problem. Relatively strong
constraints on the spectrum extension for typical Seyfert galaxies can
be only achieved through the analysis of composite spectra, i.e.
summed spectra of a number of objects. Such a composite spectrum 
was published by Gondek et al. (1996) on the basis of
combined Ginga/OSSE data for 7 Seyfert 1 galaxies. We use this
spectrum to reduce the degrees of freedom in our models.

\subsubsection{Compton amplification factor} 

A relatively low dispersion in the observed X-ray slopes of Seyfert galaxies
around the mean value 0.9 allows us to estimate the characteristic Compton 
amplification 
factor. We use for 
this purpose the Monte Carlo results of Janiuk, \. Zycki \& Czerny 
(2000).

The Comptonization process by a hot spherical cloud depends on two 
parameters: optical 
depth of the cloud, $\tau$, and electron temperature, $T_e$, and 
leads to a broad range of 
X-ray slopes. However, for each value of $T_e$ we can find an 
optical depth which 
gives a spectral slope of 0.9. Therefore we finally obtain the value 
of the Compton 
amplification factor as a function of $T_e$ only. The corresponding 
plot is shown in Fig. ~\ref{figamplif}. 

Determination of the value of $T_e$ from the observed high energy 
cutoff in the X-ray 
spectra is rather uncertain. Gondek et al. (1996) give the value of 
260 keV for the 
combined Seyfert 1 spectrum from the Ginga/OSSE data. Fortunately, the 
dependence 
of $A$ on $T_e$ is weak. This is not surprising since both the 
spectral slope and the 
Compton amplification factor are mostly determined by the value of 
the Compton 
parameter $y$.

Adopting Te = 260 keV, we fix the value of the Compton amplification factor $A = 4.0$ in 
further considerations (Fig.~\ref{figamplif}). 
It reduces the number of free parameters of our toy model to 3. 

\subsubsection{Expected trends in stationary parameters}

Fixing the value of the Compton amplification factor leads to a relation
between the original model parameters: $C$, $a$, $\beta_d$ and $\gamma$. 
It leaves a choice of 3 parameters out of 4, as basic independent 
model parameters. We analyse the relation in order to make the most convenient
choice. We  also study the dependence of
the predicted luminosity ratios on our selected set of parameters.

\medskip
\noindent{\bf Negligible dark side contribution}
\smallskip

We consider first the case of negligible contribution of the dark 
sides of the clouds to the 
UV spectrum, due to their large optical depth ($\beta_d$ = 0). 

\begin{figure}
\epsfxsize = 8.8cm \epsfbox{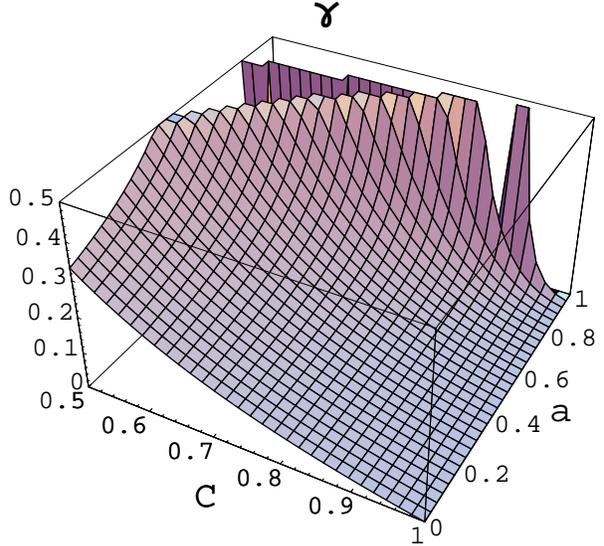}
\caption{The dependence of the probability, $\gamma$, of UV photon upscattering  
by the hot plasma on the covering factor $C$ and the X-ray albedo $a$; fixed parameters are
$\beta_d = 0$ and 
$A=4$.} \label{fig:gamma-beta0}
\end{figure}

The stationarity condition allows to express the probability of 
scattering by the hot
cloud, $\gamma$, as the function of the  X-ray albedo, $a$, and covering 
factor, $C$. We show 
this dependence in Fig.~\ref{fig:gamma-beta0}. 

Factor $\gamma$ depends mostly on the covering factor but it also shows a 
trend to increase with the albedo (i.e. the ionization parameter). 
Only large values of the 
covering factor are allowed, as the optical depth of the hot cloud is 
small and the
physical size of the X-ray region $r_X$ cannot be larger than 
$r_{UV}$ within the 
frame of our model. So $\gamma$ cannot be larger than $\tau$. It 
means that $C$ cannot be smaller than about 0.7 for $\tau = 0.1$ although larger
optical depth allow for smaller covering factors.

\medskip
\noindent{\bf The role of the dark side contribution}
\smallskip

The contribution from the dark sides of clouds may not, however, be 
negligible. 
Therefore, it is more appropriate to present the results by 
constraining the albedo. For a 
broad range of the ionization parameter $\xi$, the radiative transfer 
computations 
(Dumont  \& Abrassart 2000)  show that the X-ray albedo, $a$, for partially 
ionized gas is 
about 0.5. If we use this constraint, it leaves us a choice of two independent
parameters among $C$, $\gamma$ and $\beta_d$.

\begin{figure}
\epsfxsize=8.8cm \epsfbox{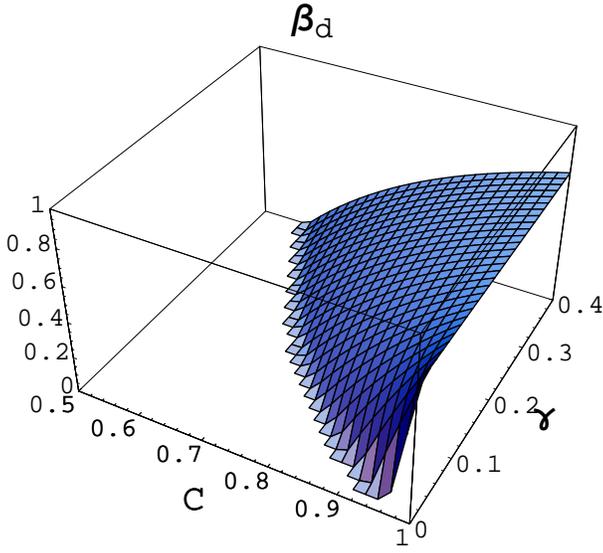}
\caption{The dependence of the fraction, $\beta_d$, of absorbed X-ray flux to
be reemitted by the dark sides of the clouds  on the covering factor,
$C$, and the probability of
scattering  by hot plasma, 
$\gamma$. Fixed parameters: $A=4$, $a=0.5$.}
\label{fig:beta-a5e-1}
\end{figure}

We plot the dependence of $\beta_d$ on the covering factor $C$ and the
scattering probability $\gamma$ in Fig.~\ref{fig:beta-a5e-1}. We see that 
$\beta_d$ should be
relatively important for large covering factors and large $\gamma$. There is 
also a range of parameters (small $C$ and $\gamma$) for which the formally
computed value of $\beta_d$ is negative (negative values are not plotted on
the figure). Using $C$ and $\gamma$ as independent variables, we have to be
aware of this unphysical range of parameters which 
depends to some extent on the adopted albedo. If we choose another set of two
arbitrary parameters the unphysical range does not disappear.

The cross-section of Fig.~\ref{fig:beta-a5e-1} for constant $\beta_d$ 
produces  
$\gamma$ 
closely correlated to $C$. Such a relation for $\beta_d=0.2$
(it is a high value for the high optical thickness of our clouds, according to the transfer
computations of Dumont \& Abrassart, 2000) is rather similar to results for  
$\beta_d = 0$. 

\subsubsection{Expected trends in the luminosity ratios}

The observed ratio of the X-ray to the UV luminosity depends only on the covering
factor if the X-ray albedo and Compton amplification factor are fixed; it is
not influenced by the second model parameter ($\gamma$ or $\beta_d$) 
within the frame of our toy
model (see Eq.~\ref{eq:lrat5}), in opposite to other quantities. 

This quantity is one of the important observables, although in practice the
determination of the $\left({L_X\over L_{UV}}\right)_{obs}$ ratio may not be
easy because of extinction. It is also of major importance in all
accretion models, so we have to study its dependence on the parameters
which we usually fix, namely Compton amplification factor $A$ and X-ray albedo $a$.

\begin{figure}
\epsfxsize=8.8cm \epsfbox{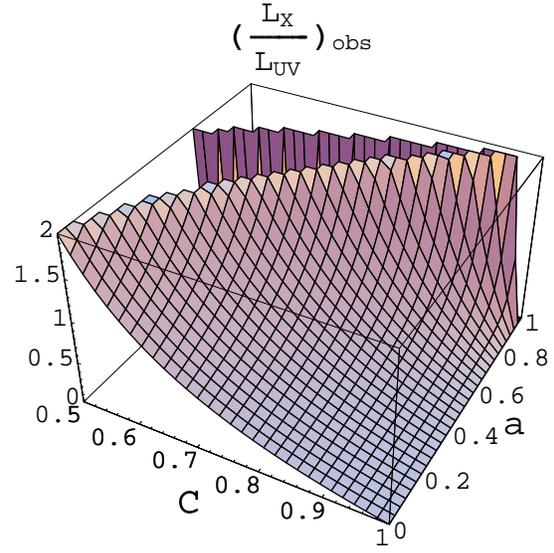} 
\caption{The dependence of the 
$\left({L_X\over L_{UV}}\right)_{obs}$ ratio on the covering factor,
 $C$, and the X-ray albedo, 
$a$. Fixed parameter: 
$A=4$.} 
\label{fig:LxsLuvobs-beta0}
\end{figure}

In Fig.~\ref{fig:LxsLuvobs-beta0} we show the $\left({L_X\over L_{UV}}\right)_{obs}$ ratio. We see that this ratio 
mostly depends on the 
covering factor.
For an albedo $a = 0.15$ (cold matter) the X-ray luminosity starts to 
dominate 
the bolometric output only if the covering factor is smaller than 
0.6. In the case of a larger albedo of partially 
ionized gas, $a=0.5$, the X-ray emission dominates the UV luminosity up to about 
$C=0.75$.

We see from this plot that $\left({L_X \over 
L_{UV}}\right)_{obs}$ is smaller than unity for the range of 
covering factors we consider. 
{\it This is a very important characteristic of this model: it is able to  account for a low X-ray to 
UV luminosity ratio without any ad-hoc hypothesis}.

\begin{figure}
\epsfxsize=8.8cm \epsfbox{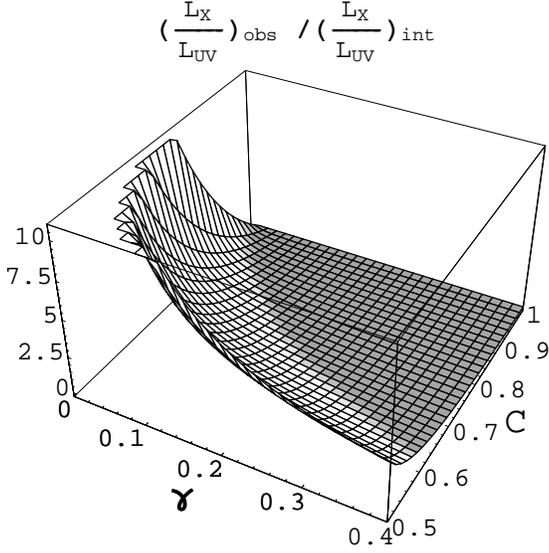}
\caption{The dependence of$\left({L_X \over 
L_{UV}}\right)_{obs} / 
\left({L_X \over L_{UV}}\right)_{int}$ on the covering factor, $C$, and the probability of
scattering  by hot plasma, 
$\gamma$. Fixed parameters: $A=4$, $a=0.5$. The non-shaded part 
corresponds to $\beta_{d} < 0$.}
\label{fig:LxsLuvobssLxsLuvi-a5e-1}
\end{figure}

Fig.~\ref{fig:LxsLuvobssLxsLuvi-a5e-1} 
 shows the dependence of  
$\left({L_X \over L_{UV}}\right)_{obs}/\left({L_X \over 
L_{UV}}\right)_{int}$ on 
the covering factor $C$ and $\gamma$ for $a=0.5$. This ratio, in contrast to
 $\left({L_X \over L_{UV}}\right)_{obs}$, depends on $\gamma$ as well, not just on the
covering factor.


Over most of the allowed parameter space, $\left({L_X \over 
L_{UV}}\right)$ is slightly lower inside the clouds system than what is observed. 
It is also interesting to note that the X-ray to UV luminosity ratio 
seen outside the cloud system can be larger than this ratio determined
inside the medium, or in other words that {\it inside the medium the radiation field is softer
than the observed spectrum}. The effect is stronger 
towards small values of $\gamma$, 
but the parameter space where this happens is rather narrow, 
because the highest values happen where there are no physical solutions.
Indeed, the energy losses through the dark sides of the clouds take 
negative values for
a large range of covering factor and probability of scattering by the hot plasma (see Fig.~\ref{fig:beta-a5e-1}).

\begin{figure}
\epsfxsize=8.8cm \epsfbox{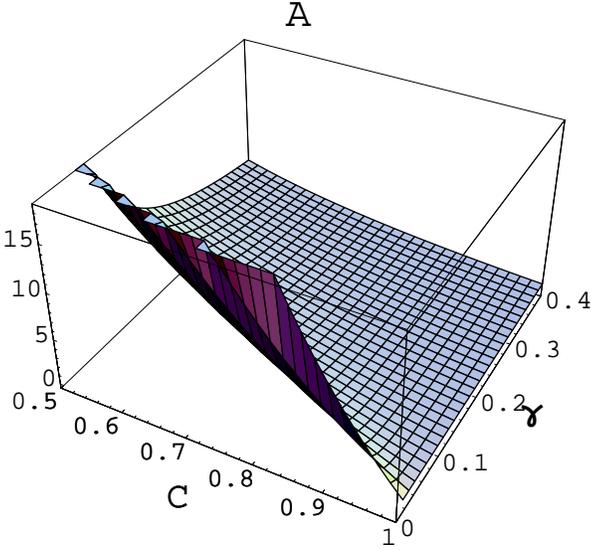} 
\caption{The dependence of the 
Compton amplification factor on the covering factor, $C$, and the 
probability of
scattering  by hot plasma, 
$\gamma$.  Fixed parameters: a=0.5, $\beta_{d}=0$.} 
\label{fig:A(Candgamma)-a0.5beta0}
\end{figure}

The reason for this can be more clearly seen on Fig.~\ref{fig:A(Candgamma)-a0.5beta0}, for which we relaxed the A=4 assumption, but fixed $\beta_{d}$. This last parameter does not qualitatively change the behavior of the equilibrium amplification factor but simply forces it to higher values when it raises. One sees that fixing A defines a unique relation in the C,$\gamma$ plane, but that there are no solutions on either side of this line. By raising $\beta_{d}$, it is possible to find an equilibrium on the side where A is too high, but the side where A is not sufficient is clearly forbidden.

\begin{figure}
\epsfxsize=8.8cm \epsfbox{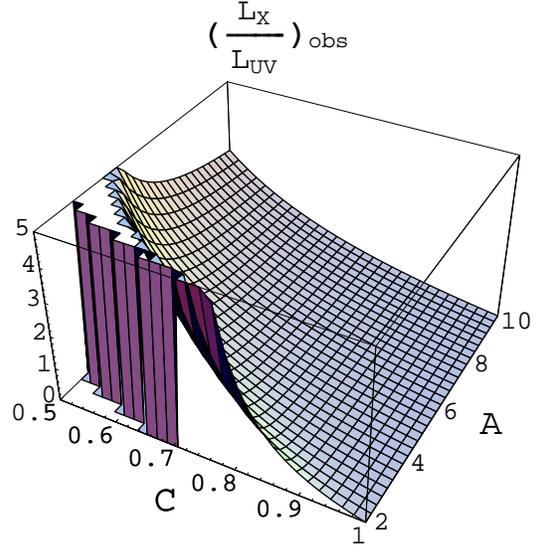} 
\caption{The dependence of the 
$\left({L_X\over L_{UV}}\right)_{obs}$ ratio on the covering factor,
 $C$, and Compton amplification factor,
$A$.  Fixed parameter: $a = 0.5$.} 
\label{fig:LxsLuvobs(CandA)-a0.5}
\end{figure}

The accuracy of the determination of the Compton amplification factor does
not influence much the X-ray to UV luminosity ratio if the covering
factor is large. 
This dependence is plotted in Fig.~\ref{fig:LxsLuvobs(CandA)-a0.5}, 
for an X-ray albedo of one half. Higher albedos (i.e. ionization state) 
require higher covering factors, whatever the amplification factor, in order to keep a reasonably low X-ray to UV luminosity ratio.

\subsubsection{Toy model for the Gondek et al. spectrum} 

We can now find model parameters which reproduce the shape of the 
mean Seyfert 
spectrum of Gondek et al. (1996). We need, for this purpose, the 
observed $({L_X \over 
L_{UV}})$ ratio, the X-ray albedo and the contribution from the dark 
sides $\beta_d$. 
Also the optical depth of the hot plasma (or its temperature) is 
required in order to obtain 
the relative size of the cloud distribution to the radius of the hot 
plasma, 
$r_X/r_{UV}$. 

We have no accurate measurements of the X-ray to the UV bolometric 
luminosity ratio for 
this sample since the UV data are very sensitive to even minor 
absorption, and the 
extension of this spectral component into the EUV is difficult to 
incorporate into the bolometric 
luminosity of the soft component. The $\nu F_{\nu}$ fluxes measured 
at 1375 \AA ~and 
2 keV given by Walter \& Fink (1993) can be used to make a rough 
estimate. Of the sources included in the combined Seyfert 1 spectrum 
by Gondek et al. 
(1996), two are heavily absorbed in the UV: MCG 8-11-11 and MCG-6-30-15 
so they have to 
be rejected. The remaining five sources have the average ratio $\nu 
F_{\nu}$ at 1375 \AA ~  to 
$\nu F_{\nu}$ at  2 keV about 14, with the highest value being 
$\sim 20$. Since the 
X-ray component is much broader, we have to apply a bolometric 
correction $K$ to it in 
order to roughly reproduce the broad band luminosity ratio. We 
assume $K=4$ so we 
obtain $(L_X/L_{UV})_{obs} = K/14=
0.3$. This is only a rough estimate; the result of detailed 
analysis would depend 
significantly on the temperature of the $UV$ component. 

Such an observed luminosity ratio can be used to estimate the 
parameters of the cloud 
distribution. Assuming the
value of albedo for partially ionized matter $a=0.5$, $A = 4$, 
$(L_X/L_{UV})= 0.3$ 
and neglecting the contribution from the dark sides ($\beta_d=0)$ we 
obtain the covering 
factor $C = 0.87$ and $\gamma$ equal 0.052 (for $a=0.3$, $\gamma$ is the 
same, and $C=0.91$). The value of $\gamma$ 
can be used to 
determine the $r_X/r_{UV}$ if we know the optical depth of the hot 
cloud. Taking the 
optical depth $\tau = 0.1$ after Gondek et al. (1996) we obtain 
$r_X/r_{UV} =0.72$. 
However, observational determination of $\tau$ strongly depends on 
the (rather poor)
accuracy of the determination of the high energy cut-off. Assuming 
a larger value of $\tau = 
0.3$ we obtain $r_X/r_{UV} =0.42$. We therefore see that we can
determine reliably the 
parameter $\gamma$, but the factorization of its value
between the physical size of the hot cloud and the optical depth is 
only weakly constrained
by the details of the spectral shape. 

Allowing for a relatively high ratio of energy leaking through the dark 
sides, $\beta_d = 0.1$, 
we obtain: $C=0.93$, $\gamma=0.03$ which translates to $r_X/r_{UV}$ 
equal 0.55 
for $\tau = 0.1$ and 0.32 for $\tau = 0.3$.

\subsection{Random variability}

Detailed observations of the variability pattern are available only 
for a few sources and the 
observed trends are possibly not characteristic of all AGN as a 
whole. Therefore we first 
outline the most basic trends and later on we apply the toy model to the 
best monitored  sources 
in order to check whether cloud obscuration may account for 
their specific 
behavior.

\subsubsection{X-ray variability amplitude} 

The X-ray variability amplitudes predicted by the
cloud model can be easily 
estimated from Eq.~\ref{xampl} in the case of no leakage from the dark side. 


A large covering
factor leads to large variability even if the number of clouds 
is large. For example, 
1000 clouds covering 0.9 of the source produce a maximum to minimum 
flux ratio 9 (see Sect.~\ref{sect:randam}). 
This factor reduces to only 1.2 for covering factor 0.5; only considering 
significantly smaller number of clouds would increase
the variability. 

The mean Seyfert 1 spectra analysis suggested that the typical 
covering factor
is about 0.9 (see Sect.~\ref{sect_mean}). Those objects display an 
X-ray variability by a 
factor 2 on average. This means that the required number of clouds 
is of the order of a few thousands.

\subsubsection{X-ray/UV relative amplitude} 

The ratio $R$ of the normalized variability amplitudes in the X-ray 
band to that in the UV band is fully determined
by the parameters of the toy model (see Eq.~\ref{eq:R}).

This ratio is equal to unity in the toy model as long as the contribution of the dark sides of the clouds 
is negligible (see Eqs.~\ref{xampl}, \ref{uvampl1}, \ref{uvampl2}). 

\begin{figure}
\epsfxsize = 8.8cm \epsfbox{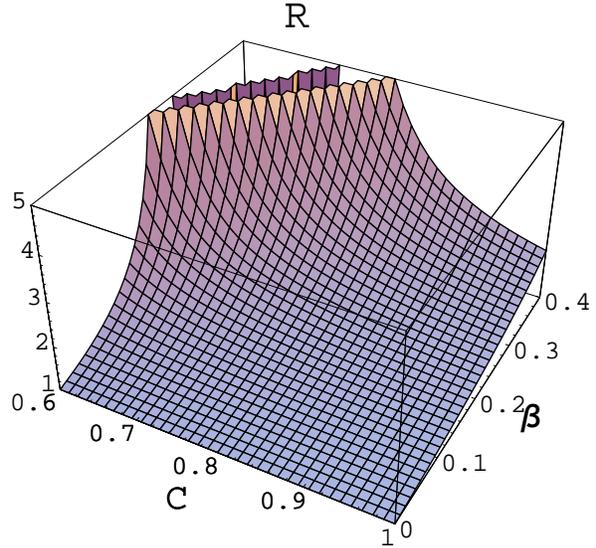}
\caption{The dependence of the ratio, $R$, (see Eq.~\ref{eq:R}) of the normalized variability 
amplitude in the X-ray and 
in the $UV$ band on $\beta_d$ and $C$. Other parameters: $a = 0.5$, $A = 
4$.}
\label{figrelamp}
\end{figure}

When the contribution from the dark sides of clouds is allowed, the 
relative amplitude in X-ray and $UV$ is reduced. We show this trend in 
Fig.~\ref{figrelamp} plotting $R$ 
against $\beta_d$ and $C$, assuming $a=0.5$, $A=4$. The value of 
$R$ is always greater than 1 and even very high values are allowed if 
$\beta_d$ is large.

\subsubsection{Application of the toy model to monitored objects}
\label{sect:moni}

A few AGN were recently monitored both in the UV and X-ray bands, and
the results of these campaigns can be used directly to determine
the properties of the cloud distribution in those sources within the
frame of our random variability picture.

Observational data provides us either directly with the rms values both
in UV and X-rays or with the value of variability factor $\cal A$ in
those bands. Data were taken from Goad et al. (1999) and Edelson \& Nandra 
(1999) for NGC 3516, from Nandra et al. (1998) for NGC 7469, from
Edelson et al. (1996) for NGC 4151 and from Clavel et al. (1992)
for NGC 5548.

Observational estimation of the ratio of the X-ray to UV luminosity is
unfortunately rather complex. Most sources have red optical/UV
spectra, with negative slopes on a $\nu F_{\nu}$ plot. It most probably means that
the objects are considerably reddened, with a significant fraction of
the energy reemitted in the IR (e.g. Wilkes et al. 1999). 
A conservative discussion of this problem
for NGC 4151 by Edelson et al. (1996) concluded
that X-ray luminosity is 3 times smaller than the 
UV/optical/IR luminosity. The situation for the other objects looks qualitatively similar.
Also this ratio determined for the Gondek et al. (1996) composite is of the 
same order (i.e. 0.3). We therefore adopted $L_X/L_{UV}$ ratio equal
1/3 for all objects. 

We now fix two of the toy model parameters: the Compton amplification factor
$A=4$ and X-ray albedo $a = 0.5$. Now, two observed quantities
(normalized variability amplitudes in the UV and X-ray bands)  allow us
to calculate the covering factor $C$, the number of clouds, $N$,
the probability of upscatering for a UV photon $\gamma$, and the
contribution of the dark sides of clouds, $\beta_d$. The results
for the four selected objects are given in Table~\ref{tab:objects}.

The covering factor obtained is the same for all objects, as it is determined
purely by the X-ray to UV luminosity ratio.
The assumption of $L_X/L_{UV}=1/3$ influences the obtained value
of the covering factor, but not strongly. A value of 0.5 would give 
$C=0.86$ and
a value of 0.1 would give $C = 0.96$. The last 
value is more appropriate for quasars than for Seyfert 1 galaxies.
Also the adopted value of the albedo does not influence the results significantly:
$a=0.3$ would give a covering factor of 0.86 for our objects.

The value of $\beta_d$ is equal to zero in two of our objects,  since 
the normalized variability in the UV and X-rays
are equal and no contribution from the dark sides is required.

  \begin{table}
  \caption{Toy model parameters for AGN.
  \label{tab:objects}}
  \begin{tabular}{rrrrrrr}
   \hline
      &             &           &   &   &        &        \\ 
Object & $rms_{UV}$ & $rms_{X}$ & C & N &$\gamma$& $\beta_d$ \\
 NGC  &   observed  &  observed  &         &   &    &  \\
      &    &  &                  \\
\hline
      &     \\
 3516 & 0.333 & 0.357 &  0.90 &  1180 &  0.047  &  0.04 \\
 7469 & 0.167 & 0.167 &  0.90 &  5400 &  0.044  &  0.00 \\    
 4151 & 0.009 & 0.024 &  0.90 &  2610 &  0.095  &  0.39 \\
 5548 & 0.222 & 0.222 &  0.90 &  3050 &  0.044  &  0.00 \\

  \end{tabular}
  \end{table}

\subsection{Mean spectrum from radiative transfer codes} 

Since the complete solution of radiative transfer equation within the cloud
system are extremely time consuming, we used the toy model results to
guide our choice of the model parameters. Now we can test our toy model
parameterization against numerical results.

The present version of the radiative transfer computations does not include yet
the finite size of the hot Comptonizing medium located close to the black
hole, so the effect of Comptonization is replaced by a central
point source emitting a power law continuum of an arbitrary slope and
extension.

\begin{figure}
\epsfxsize=8.8cm \epsfbox{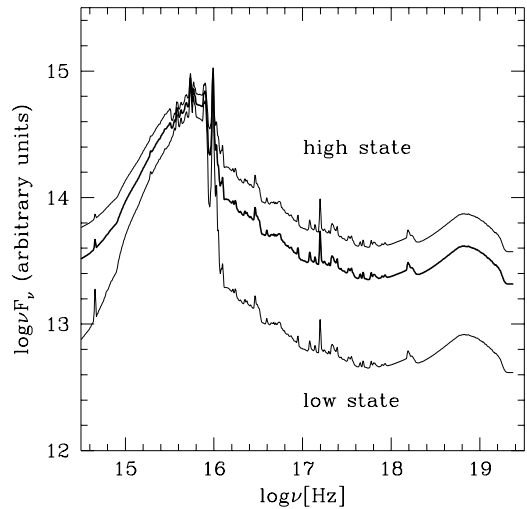} 
\caption{The mean spectrum (thick line) calculated with the coupled codes 
{\sc{titan}} and 
{\sc{noar}} for the following parameters of the shell: log$ n =14$, log $N_H =26$, 
covering factor $C =0.9$. The incident primary radiation was assumed to be
a power law extending from 1 eV to 100 keV, with energy index $\alpha = 1$,
and ionization parameter $\xi=300$. The size of the central source was
neglected. The two other spectra (thin lines) show the high and low state
expected as variations at a 2 standard deviation level in the case of 
$N=1000$ clouds (see Sect.~\ref{sect:randam}).} 
\label{fig:aam4}
\end{figure}

The mean spectrum from our computation is shown 
in Fig.~\ref{fig:aam4}, along
with the parameters used.

Spectral features in the UV and X-ray band are clearly visible, including 
iron $K_{\alpha}$
line and a number of emission lines in soft X-rays. These features are due to
'reflection' of X-rays by partially ionized clouds.

\begin{figure}
\epsfxsize=8.8cm \epsfbox{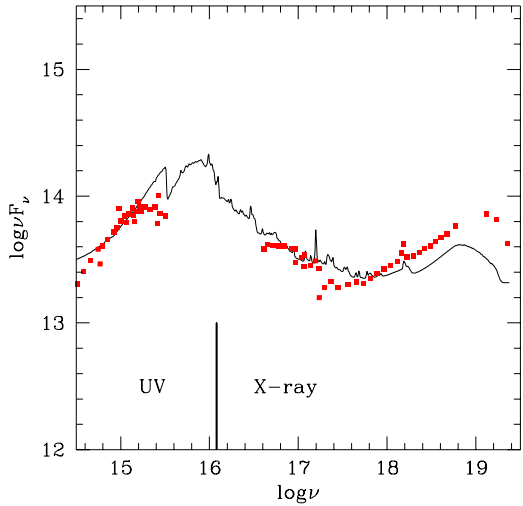} 
\caption{The mean spectrum calculated with the coupled codes {\sc{titan}} and 
{\sc{noar}} for the parameters  as in Fig.~\ref{fig:aam4}, but with the contribution
from the dark sides neglected, as required by the variability analysis of
the Seyfert 1 galaxy NGC 5548 (see Table~\ref{tab:objects}). Filled squares 
show 
schematically the observed spectrum of NGC 5548 after Magdziarz et al. (1998).} 
\label{fig:aam}
\end{figure}

We can now estimate from the computations some of the toy model parameters.

The ratio of the X-ray to UV luminosity can be now calculated from the model,
assuming a certain division between these two energy bands. Looking at the
Fig.~\ref{fig:aam4} we clearly see the 
two-component character of the spectrum,
with a big blue bump extending roughly to 50 eV, and an X-ray component of 
roughly a power law shape. Therefore we adopt 50 eV
as a division point between the UV and X-ray bands.

The observed $L_X/L_{UV}$ ratio depends on the adopted size of the central
cloud. If this size is neglected, as in this computation, the  
$\left({L_X\over L_{UV}}\right)_{obs}$ is equal to 0.24. The assumption of 
 $r_X/r_{UV}=0.7$
reduces it to 0.22, so the effect of this uncertainty on the spectral
appearance is not essential.

The frequency-averaged albedo in the full numerical computations, calculated as
a ratio of the X-ray luminosity to the total incident luminosity, is 
equal to 0.58, in 
agreement with the values used to construct 
Fig.~\ref{fig:LxsLuvobs-beta0}-\ref{fig:A(Candgamma)-a0.5beta0}.

We can also calculate the effect of leaking through the dark sides of the
clouds. For the presented computations $\beta_d = 0.26$.

This value is much higher than in three out of the four monitored objects. It 
means that in Seyfert galaxies like NGC 5548, the column density of
clouds should be larger than $N_H=10^{26}$ cm$^{-2}$ adopted in the
computations. Therefore we compare the mean observed spectrum of NGC 5548
with the computed one, neglecting the contribution from the dark side of clouds,
i.e. $\beta_d = 0$ (see Fig.~\ref{fig:aam}). The predicted spectrum may still
be too bright in far UV range, which suggests that the ionization
parameter adopted in the numerical computations was slightly too 
low. Also,
the high energy cut-off adopted in the computations (100 keV) should be
increased in order to fit this particular data. When the
dark side contribution is neglected, the obtained 
$\left({L_X\over L_{UV}}\right)_{obs}$ ratio is higher, equal to 0.43.

From the toy model, if we assume $C=0.9$,  $a = 0.58$ and $A=4$ 
we obtain $\left({L_X\over L_{UV}}\right)_{obs} = 0.38$. Therefore, the
toy model reproduces surprisingly well the complex solutions of the
radiative transfer if supplemented with appropriate values for the X-ray
albedo and dark side emission efficiency.

\subsection{Random variations from the radiative transfer codes}

Random rearrangement of the clouds leads to minor changes of the effective
covering factor in the direction towards the observer. We illustrate here
these changes using the results of the numerical computations described in
detail in the previous section.

Assuming the number of clouds, $N=1000$, we show two representative examples 
of the observed spectra at any moment of time (see Fig.~\ref{fig:aam4}).
Since the contribution of the dark sides was non-negligible for the adopted
parameter set, the variations seen in UV are of much lower amplitude than the
variations of the X-ray emission.

\section{Discussion}

\subsection{Quasars, Seyfert 1 galaxies and Narrow Line Seyfert 1 galaxies}

Various classes of AGN differ systematically although it may not necessarily 
mean that they form truly separate classes. Instead, they rather represent
various parts of the same continuous multidimensional distribution in
some parameter space. Interesting approach to this problem was formulated 
within the frame of the method of principal component analysis (PCA) by
Boroson \& Green (1992) and Brandt \& Boller (1998).

Our cloud model can well reproduce the typical trends.
Stronger Big Blue Bump component characteristic of quasars and 
narrow line Seyfert 1 galaxies (NLS1)
corresponds to a larger covering factor, of order of 0.98, instead of
0.9 obtained for Seyferts (see Sect.~\ref{sect:moni}). 
Stronger variability of NLSy1 galaxies in 
comparison with quasars (see Leighly 1999) means that the number of clouds in these objects
is of order of $10^4$, not much higher than for Seyfert 1 galaxies but
an order of magnitude lower than in quasars.

These trends are not surprising. If we imagine for simplicity that all
accreting clouds are identical, the number of clouds present would depend
on the accretion rate and the travel time of clouds scales with the
mass of the black hole, $N \propto \dot M M$. The covering factor would
be given by the number of clouds and the size of the typical radius
$r_{UV}$, scaling again with the mass, i.e. 
$C \propto N/M^2 \propto \dot m/M$. This would mean  that the covering factor
$C$ is mostly determined by the luminosity  to the Eddington luminosity ratio
and both quantities are high in quasars and NLSy1 galaxies, while additionally the
number of clouds depends on the mass and accretion rate, causing fainter
objects to vary more than brighter objects with the same luminosity ratio.  

\subsection{Relative normalized amplitudes in X-rays and UV}

Our model of variability predicts that the normalized
amplitude of  variations in X-rays should be equal to  that in the UV
if the contribution of the dark sides of the clouds to the UV emission can be
neglected. Therefore, equality of these two amplitudes in a number
of objects is expected. In those objects which show stronger variability
in the X-ray band than in the UV band, a certain level of contribution from the
dark sides is required.

This kind of behavior is quite in contrast with the frequently 
adopted picture in which 
the UV variability is caused by reprocessing of the variable X-ray 
flux. In this case the 
basic effect is in the change of the temperature of the illuminated 
cool gas. Equality of the normalized variability amplitudes in X-rays
and in the UV requires fine tuning. We can 
estimate the effect in the following way.

\begin{figure}
\epsfxsize = 8.8cm \epsfbox{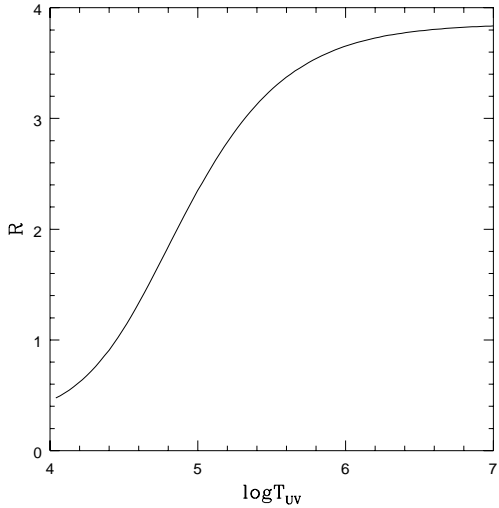}
\caption{The ratio $R$ (see Eq.~\ref{eq:R}) of the normalized variability amplitude at 1315 
\AA and 2 keV
as a function of cold matter temperature within the frame of 
reprocessing scenario of 
$UV$ variability. We assumed factor 2 variability of the
variable irradiating X-ray flux}
\label{figbbrep}
\end{figure}

In Fig.~\ref{figbbrep} we show the result of complete thermalization 
of the X-ray
flux varying by a factor 2, described using a simple black body 
approach. The variability 
in the UV (1315 \AA) depends on the mean value of the cold matter 
temperature $T_{cl}$. 
$R(T_{cl})$ is close to 1 only for a single specific value of the 
mean cloud temperature 
($\sim 27 000$ K).
For colder matter, the UV variability is stronger since we are at the 
exponential
part of the Planck function. Hotter clouds exhibit variations with an
amplitude four times smaller than the amplitude in the X-rays which caused variations.

Fine resolution monitoring of the Seyfert galaxy NGC 7469 in the UV and 
X-ray bands
show that, at least in this source, the relative amplitudes in the UV and X-ray 
bands are the same. It is easily explained within the frame of the cloud model 
by adopting a relatively low contribution of the dark sides of the clouds to the UV 
band, i.e. rather high column density of the clouds. In the case of the
simple reprocessing picture, the same value for both normalized amplitudes is
difficult to obtain (see Fig.~\ref{figbbrep}). The typical behavior of 
the other three sources considered in Sect.~\ref{sect:moni} also conforms to 
the cloud model expectations, with one more source showing the same amplitude
ratio (NGC 5548). Unfortunately, the number of objects monitored both in the UV 
and X-rays is
still low.

Within the frame of our model,  large amplitude of variations should be 
unavoidably accompanied by 
occasional complete coverage of the X-ray source due to large 
statistical 
deviations. Such events are actually observed; the best documented case was 
seen 
in NGC4051 (Guainazzi et al. 1998, Utley et al. 1999).

Surprisingly low amplitude in the optical band, or no variations seen in 
some sources strongly
variable in X-ray band (see Miller 1999,: also Boller, privite communication), 
can be also accommodated 
by our model as it may 
correspond to a very large contribution of the dark sides of clouds. However, 
this
effect may be also due to the domination of the optical band by the starlight
of the host galaxy.

Our obscuration model of variability cannot explain 
those events which are characterized by larger amplitude variations in the UV
than in the X-ray band. Such events are occasionally observed
in some sources. For example, NGC 5548 showed a spectacular
brightening in the UV in 1984 which was accompanied by a very moderate
brightening in X-ray band (see Clavel et al. 1992). The nature of
such long timescale events should be definitively different
from random variability of the covering factor  discussed in
this paper.

\subsection{Hard X-ray slope and high energy cut-off}

The prediction of the obscuration model of variability is that nothing
actually changes within the source. Random rearrangements of clouds
blocking our line of sight to the X-ray emitting region do not change
either the optical depth, or the temperature of the hot plasma.
The amount of reflection also should in principle not vary with respect to 
the primary emission.  

Observational constraints on the variability of the slope of the X-ray
'primary' component are not conclusive. A number of studies suggest that
strong variations in the luminosity are not accompanied by spectral
changes (e.g. Turner, George \& Netzer 1999 for Akn 564, George et al. 1998
for NGC 3227; see also Gierli\' nski et al. 1997 for Cyg X-1). 
However, Zdziarski, Lubi\' nski \& Smith (1999) and 
Done, Madejski \& \. Zycki (1999) suggest the presence
of a correlation between the slope of the primary component, amount of reflection
and the source luminosity for NGC 5548. The problem of decomposition of the
spectrum into two components - primary and reflected - is difficult for the 
short time sequences of the data used in variability studies. Also, 
a fraction of the reflection comes from very large distances and does not 
respond to the nuclear emission within the observed time, as suggested by 
the remnant emission observed during the off state in NGC 4051 
(Guainazzi et al. 1998)

The prediction of the model that no spectral variability in X-rays is expected
is also not firm. In a real situation, if there are any temperature
gradients within the hot cloud and the cold clouds are located on a range
of distances with a complex overlapping pattern, we can expect some weak
spectral variations caused by cloud eclipses. However, the predictions of
the trends would require either an ad hoc parameterization, or a dynamical 
study of cloud formation and disruption which is beyond the scope of the 
present paper. 


\section{Conclusion}

The cloud scenario offers an interesting quasi-spherical model of accretion
onto a central massive black hole. It explains the observed large ratio
of the Big Blue Bump luminosity to the X-ray luminosity. 
The second prediction
of the model is the variable obscuration of the X-ray source. It
can explain the following observed
trends:

\begin{itemize}
\item{$\clubsuit$} significant X-ray variability not accompanied by the change of the
spectral slope; this is due to random cloud redistribution without changes of the covering factor,
\item{$\clubsuit$} amplitude of the variability in X-ray band larger or equal
to amplitude  in UV
band: this is due to the contribution from the dark sides of clouds,


\item {$\clubsuit$} variability timescales ranging from $10^2$ s to $10^6$ s; 
this is due to the size
of the optically thick clumps, the size of the X-ray medium and Keplerian motion.
  
\end{itemize}

\noindent Variability analysis
indicates that in many objects the column density of the clouds is very large,
$N_H >>10^{26}$cm$^{-2}$ suggesting their origin in violent disk disruptions.

Further progress is required in order to incorporate the finite size of the
hot central cloud into the numerical computations of the radiative transfer,
and to address the problem of intrinsic variability.

\begin{acknowledgements}
We are grateful to Suzy Collin for many extensive discussions and suggestions,
to Anne-Marie
Dumont for participation in computing the numerical model and for helpful 
comments to the manuscript, and to Katrina Exter for her help with English. 
We also thank Dirk Grupe, our referee, for his help in improving the presentation of
the paper and for pointing out valuable references.
Part of this work was supported by grant 2P03D01816 of the Polish 
State Committee for 
Scientific Research and by Jumelage/CNRS No. 16 ``Astronomie 
France/Pologne''.
\end{acknowledgements}

\section*{Appendix A : Energy conservation and Compton amplification factor}

We denote the number of UV photons inside the radius $r_{UV}$ by 
$N_{UV}$, the 
number of X-ray photons by $N_X$, and their mean energies by 
$E_{UV}$ and 
$E_X$, correspondingly. We also introduce the efficiency $\eta_X$ of 
creating an X-ray 
photon of an energy $E_X$ during subsequent scatterings within
the hot plasma and the efficiency $\eta_{UV}$ of creating UV photons 
from an absorbed
X-ray photon of energy $E_X$. The value of $E_X$ and $\eta_X$ depend 
sensitively on 
the
hot cloud optical depth $\tau$ and its electron temperature $T_e$ 
since they are
related to the spectral shape of produced X-rays. $\eta_{UV}$ is 
simply given by
\begin{equation}
\eta_{UV} = E_X/E_{UV}.
\label{PUV}
\end{equation}

The conservation law for the number of UV photons within the system 
is given by 
\begin{equation}
(P_{esc}^{UV} + P_{ups}^{UV})N_{UV} =
\eta_{UV} P_{abs}^X N_X
\label{statUV}
\end{equation}
where all the probabilities are given by Eqs. (1) - (7). 

The conservation law for a number of X-ray photons is given by 
\begin{equation}
(P_{esc}^X + P_{abs}^X+P_{dark}^{UV})N_X = \eta_X P_{ups}^{UV} N_{UV} 
\end{equation}

Those two equations can be combined into the condition of the 
stationarity 
\begin{equation}
(P_{esc}^{UV} + P_{ups}^{UV})(P_{esc}^X + P_{abs}^X+P_{dark}^{UV})= \eta_X 
\eta_{UV} 
P_{ups}^{UV}P_{abs}^X
\label{statA}
\end{equation}

If true losses from the system (i.e. leak through the dark sides and 
escape) are equal zero, 
$\eta_{UV} \eta_X$ must be equal 1, i.e. no Compton amplification is
possible (or required) and the hot and cold plasma achieve thermal 
equilibrium. Only 
stationary losses and energy supply to the hot plasma support the 
existence
of the two strongly different media.

The product $\eta_{UV} \eta_X$ is directly related to the Compton 
amplification factor 
$A$
of the hot plasma
\begin{equation}
\eta_{UV} \eta_X = A(\tau, T_E)
\end{equation}

Substituting this relation into Eq.~\ref{statA} we obtain the 
stationarity condition 
\ref{stat}.

We can also determine the ratio of the bolometric luminosity in X-
ray and UV spectral 
components.
\begin{equation}
\left({L_X \over L_{UV}}\right)_{int} = {N_X E_X \over N_{UV} 
E_{UV}}. 
\end{equation}
The $N_X/N_{UV}$ ratio can be determined from Eq.~\ref{statUV} 
\begin{equation}
\left({L_X \over L_{UV}}\right)_{int} ={(P_{esc}^{UV} + 
P_{ups}^{UV}) \over 
P_{abs}^X (1 - \beta_d)} {E_X \over E_{UV} \eta_{UV}} \end{equation}
where the last term is equal 1 on the basis of Eq.~\ref{PUV}.

\section*{Appendix B : Mean number of clouds on the line of sight}

	We assume that $N$ clouds of radius $r_{cl}$ are 
homogeneously distributed in a shell of radius $r_{UV}$ whose thickness 
is small with respect to its radius. The coverage factor of the 
system of clouds is $\Omega/4\pi=C$. The mean number of clouds on the 
line of sight, $N_{ls}$, is given by:

\begin{equation}
N_{ls} \sim N {X\over 4C}
\label{eq-Nlos}
\end{equation}

where $X=(r_{cl}/R)^2$.

	The total number of clouds is related to the coverage factor by:

\begin{equation}
X = 1\ -\ (1\ -\ C)^{1/N}
\label{eq-Nlos-bis}
\end{equation}

	This expression can be expanded to the first order:

\begin{equation}
  X={1\over N}Ln{1\over 1-C},
\label{eq-Nlos-ter}
\end{equation}

provided that $Ln{1\over  1-C}\ll N$. This is easily achieved except 
for extremely small values of $1-C$ (in our case $1-C\sim 0.1$). One 
deduces thus that $X\sim 2/N$, and therefore $N_{ls}\sim 1/C$.

\end{document}